# Direct Visualization of an Incommensurate Unidirectional Charge Density Wave in La$_4$Ni$_3$O$_{10}$

Mingzhe Li[1], Jiashuo Gong[1], Yinghao Zhu[2], Ziyuan Chen[1], Jiakang Zhang[1], Enkang Zhang[2], Yuanji Li[1], Ruotong Yin[1], Shiyuan Wang[1], Jun Zhao[2,*], Dong-Lai Feng[1,3,4], Zengyi Du[3,*], Ya-Jun Yan[1,3,*]

[1]Hefei National Research Center for Physical Sciences at the Microscale and Department of Physics, University of Science and Technology of China, Hefei, 230026, China

[2]State Key Laboratory of Surface Physics and Department of Physics, Fudan University, Shanghai, 200433, China

[3] Hefei National Laboratory, University of Science and Technology of China, Hefei, 230026, China

[4] National Synchrotron Radiation Laboratory, School of Nuclear Science and Technology, and New Cornerstone Science Laboratory, University of Science and Technology of China, Hefei, 230026, China

**Corresponding authors:** zhaoj@fudan.edu.cn; duzengyi@ustc.edu.cn; yanyj87@ustc.edu.cn

**Superconductivity emerges in both La$_3$Ni$_2$O$_7$ and La$_4$Ni$_3$O$_{10}$ under high pressure by suppressing their density-wave transitions, but critical temperature ($T_c$) differs significantly between these two compounds. To gain deeper insights into the distinct superconducting states, it is essential to unravel the nature of the density-wave states at ambient pressure, a topic that remains largely unexplored. Here, using scanning tunneling microscopy/spectroscopy (STM/STS), we report the direct visualization of an incommensurate unidirectional charge density wave (CDW) in La$_4$Ni$_3$O$_{10}$ in real space. The density of states (DOS) is strongly depleted near $E_F$, indicating the opening of a CDW gap of $2\Delta \approx 71$ meV, which is unfavorable for the formation of superconductivity at ambient pressure. We propose that the CDW arises from Fermi surface nesting, and is likely a subsidiary phase of a spin density wave. Compared to La$_3$Ni$_2$O$_7$, the weaker electronic correlation in La$_4$Ni$_3$O$_{10}$ is likely one reason for the lower $T_c$.**

The discovery of superconductivity in pressurized Ruddlesden-Popper (RP) phase La$_3$Ni$_2$O$_7$ and La$_4$Ni$_3$O$_{10}$ or their thin film form has significantly advanced the research on nickelate superconductors [1-15]. Unlike cuprate and iron-based superconductors where superconductivity exists across various structural types [16,17], so far superconductivity has only been observed in two RP phase nickelates and infinite layer thin films [1-15,18-23]. The critical temperature ($T_c$) in bilayer La$_3$Ni$_2$O$_7$ is ~ 80 K, significantly higher than $T_c$ ~ 30 K in trilayer La$_4$Ni$_3$O$_{10}$, contrasting with cuprates where the highest $T_c$ is found in systems with three CuO$_2$ layers [16]. To gain deeper insights into nickelate superconductors, comparative study of La$_3$Ni$_2$O$_7$ and La$_4$Ni$_3$O$_{10}$ is essential.

At ambient pressure, both La$_3$Ni$_2$O$_7$ and La$_4$Ni$_3$O$_{10}$ exhibit density-wave (DW) transitions around 100 K ~ 150 K [24-42], which are suppressed by high pressure and then superconductivity emerges [1-12], analogous to cuprate and iron-based superconductors [16,17]. The DW fluctuations are considered as the pairing glue of high-temperature superconductivity, making the understanding of these DWs crucial for uncovering high-temperature superconducting mechanism. For La$_3$Ni$_2$O$_7$, well-defined optical-like magnetic excitations have been observed via resonant inelastic x-ray scattering and neutron scattering [32,33], and the magnetism has been further confirmed by nuclear magnetic resonance (NMR) and muon spin relaxation studies [34,38,39]; however, the existence of charge density wave (CDW) is controversial [34,40,41], probably due to weak CDW amplitude. The existence of oxygen vacancies, structural intergrowth and phase separation in $La_3Ni_2O_7$ single crystals further hinders the investigation [3,43,44]. In contrast, $La_4Ni_3O_{10}$ single crystals show higher uniformity and sample quality. Intertwined CDW and spin density wave (SDW) were reported by x-ray diffraction (XRD) and neutron scattering studies [27], and density functional theory shows that the susceptibility reaches maxima near the SDW wave vector, indicating its origin from Fermi surface nesting [27]. Subsequently, angular resolved photoemission spectroscopy (ARPES), optical spectroscopy and NMR measurements have revealed possible DW gaps, but the gap size and momentum location vary significantly [25,35-38]. Therefore, more experimental evidence for the DWs in RP phase nickelates is needed, especially the direct visualization of their spatial distribution in real space, the exact gap size and the underlying mechanism. Scanning tunneling microscopy/spectroscopy (STM/STS), with unique high spatial and energy resolution, plays a crucial role in revealing the nature of DWs and their influence on electronic structure [45,46]. In this letter, by using STM/STS, we directly observe an incommensurate unidirectional CDW in $La_4Ni_3O_{10}$ in real space, and the density of states (DOS) is significantly depleted between -32 meV and 39 meV. Possible Fermi surface nesting scenarios are discussed, suggesting that the observed CDW could be a subsidiary phase of a SDW with $\mathbf{q}_{SDW} = 1/2\mathbf{q}_{CDW}$.

Figure 1(a) shows the crystal structure of $La_4Ni_3O_{10}$ at ambient pressure, an orthorhombic in-plane unit cell is formed due to the tilt of $NiO_6$ octahedra in *bc* plane [Fig. 1(b)]. Temperature-dependent resistance curve reveals a metal-to-metal transition at $T_{DW} \sim 138$ K [Fig. 1(c)], which was considered as concomitant SDW/CDW transitions [24,26-30,35-38,42]. After cleaving $La_4Ni_3O_{10}$ crystals (see more experimental methods in section 1 of Supplementary Material (SM) [47]), both LaO-I and LaO-II surfaces are exposed, and our STM study mainly focuses on LaO-I surface as it is atomically flat (please see section 2 of SM for more details [47]).

Figures 1(d) and 1(e) show the typical topographic image and fast Fourier transformation (FFT) image of LaO-I surface, respectively. Five sets of nondispersive diffraction spots are identified and labeled as $\mathbf{q}_{Bragg}$, $\mathbf{q}_b$, $\mathbf{q}_1(\mathbf{q}_2)$, $\mathbf{q}_3$, and $\mathbf{q}_4$, respectively. Besides $\mathbf{q}_{Bragg}$ and $\mathbf{q}_b$ that correspond to the original La atomic lattice and its $\sqrt{2}R45°$ reconstruction due to the tilt of $NiO_6$ octahedra (magenta and black boxes in Fig. 1(b)), the remaining three sets of diffraction spots cannot be accounted by bulk crystal structure [9,24,26-30]. From atomically resolved topographic image [Fig. 1(f)], positions of La atoms are identified, proving further distortion. Figure 1(g) shows the schematic lattice distortion, explaining well the additional diffraction spots and resulting in larger 2*b* and 4*b* periods in *b*-axis (please see section 3 of SM for more details [47]), which modulate the DOS simultaneously and will be discussed later. Considering that previous XRD measurements on bulk $La_4Ni_3O_{10}$ did not detect such lattice distortion [26-30] and STM is a surface-sensitive technique, such lattice distortion should exist solely on the cleaved surface.

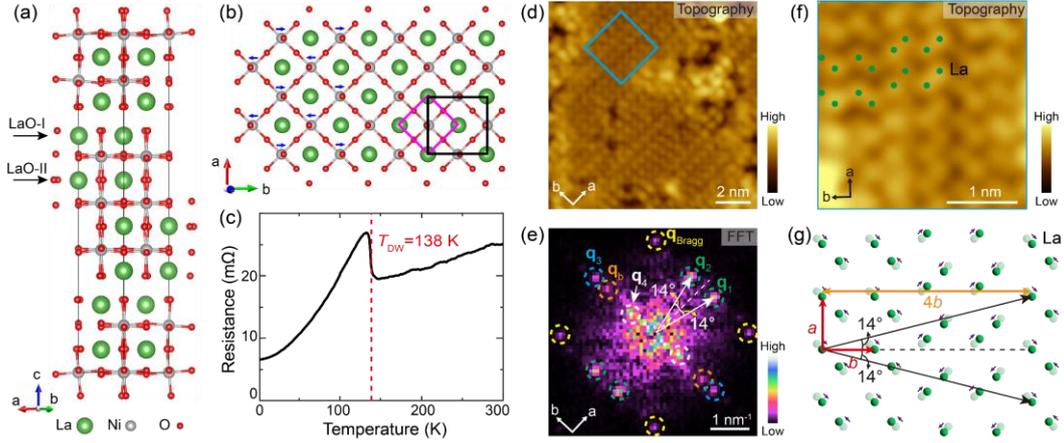

FIG. 1. Bulk crystal structure and surface lattice distortion of $La_4Ni_3O_{10}$. (a) Bulk crystal structure of $La_4Ni_3O_{10}$. (b) Top view of panel (a). The blue arrows mark the tilt directions of $NiO_6$ octahedra in *bc* plane, resulting in an orthorhombic unit cell indicated by the black box. The magenta box indicates the smallest period of La atomic lattice. (c) Temperature-dependent resistance curve of $La_4Ni_3O_{10}$. (d),(e) Typical topographic and FFT images of LaO-I surface. (f) Magnified view of the area indicated by the blue box in panel (d). The sites of La atoms are marked out by the green spots. (g) Sketch of lattice distortion on LaO-I surface. The translucent green spots represent the original La sites, which move along the violet arrows, resulting in distorted lattice as indicated by the solid green spots. A larger 4*b* period along the *b*-axis is induced, and the lattice planes correspond to Bragg spots of $q_1$ and $q_2$ are indicated by the black arrows. Measurement conditions: (d) $V_b$ = -10 mV, $I_t$ = 30 pA; (f) $V_b$ = -40 mV, $I_t$ = 100 pA.

Since the topographic images of LaO-I surface are significantly influenced by lattice reconstruction, it is difficult to discern CDW modulations; comparatively, differential conductance maps are more sensitive to charge modulations. Several representative d$I$/d$V$ maps on LaO-I surface are listed in Fig. 2(a), distinct unidirectional charge stripes along *a*-axis are observed; in the corresponding FFT images [Fig. 2(b)], a new set of diffraction spots appears along *b*-axis as indicated by cyan arrows, in addition to the diffraction spots originating from lattice reconstruction. Figure 2(c) shows FFT intensity profiles taken along cut #1 in Fig. 2(b), the new set of diffraction spots are significant and nondispersive at all measured energies; its wave vector $q_{CDW} \approx 0.76 q_b$, consistent with previous XRD report [27]. Therefore, it is tentatively assigned as CDW, and we will provide more evidence below and investigate whether a CDW gap opens at $E_F$.

Figure 3(a) shows a typical topographic image of the same sample region as in Fig. 1(d), and Fig. 3(b) displays the typical d$I$/d$V$ spectrum collected on it. The DOS is strongly depleted within approximately ±40 meV, resulting in a roughly symmetric gap-like feature at $E_F$. Additionally, there are several distinct peaks located at approximately -130, -44, -32, 39 and 84 meV, which are labelled as $P_1 - P_5$, respectively. Similar spectrum is observed on LaO-II surface, except that the sharp peaks and gap-like feature are slightly weakened (section 4 of SM [47]). Figures 3(c) and 3(d) display the spatial DOS oscillations along cuts #2 and #3 in Fig. 3(a). The DOS is distributed uniformly along *a*-axis [Fig. 3(c)], while it is strongly modulated along *b*-axis [Fig. 3(d)]. Taking $P_1 - P_5$ as examples, Fig. 3(e) illustrates the spatial DOS oscillations at corresponding energies, revealing significantly different periods.

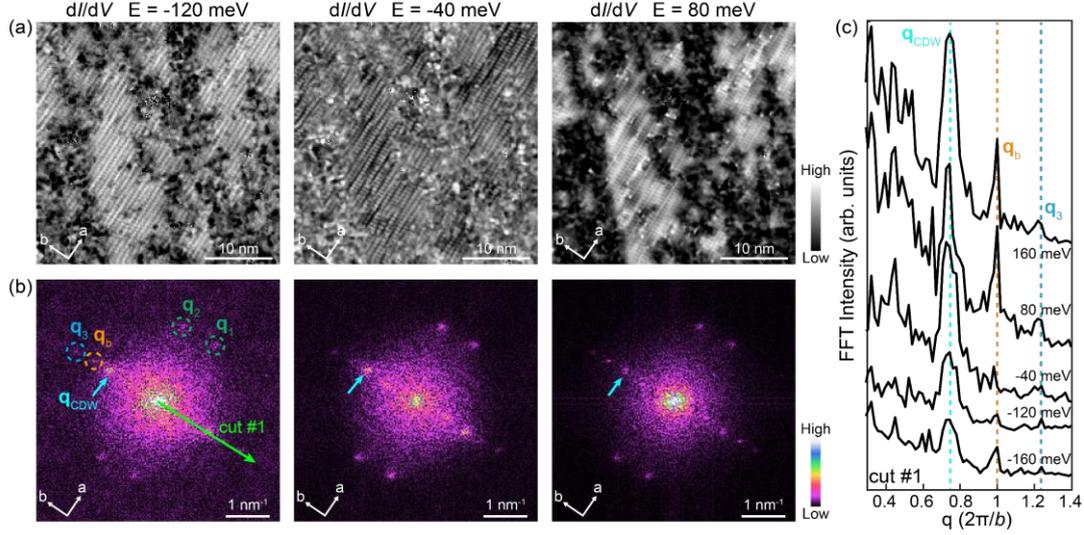

FIG. 2. Direct visualization of CDW on LaO-I surface. (a),(b) Representative d$I$/d$V$ maps and corresponding FFT images under different energies. (c) Profiles of FFT intensity at different energies, taken along cut #1 in panel (b). Measurement conditions: (a) $V_b$ = -200 mV, $I_t$ = 50 pA, $\Delta V$ = 14 mV.

As discussed above, lattice reconstructions and CDW coexist along $b$-axis and modulate the DOS simultaneously. Therefore, we consider four cosine components with periods of $b$, 1.316$b$, 2$b$ and 4$b$ to fit the experimental data shown in Fig. 3(e), here the 1.316$b$ period corresponds to $\mathbf{q}_{CDW}$. The fitting function is expressed as follows:

$$\frac{dI}{dV}(d) = A_b \cos\left(2\pi \frac{d}{b} + \varphi_b\right) + A_{CDW} \cos\left(2\pi \frac{d}{1.316b} + \varphi_{CDW}\right) + A_{2b} \cos\left(2\pi \frac{d}{2b} + \varphi_{2b}\right)$$
$$+ A_{4b} \cos\left(2\pi \frac{d}{4b} + \varphi_{4b}\right) + C_1 d + C_2$$

where $b$, d$I$/d$V$, and $d$ represent the $b$-axis lattice constant of La$_4$Ni$_3$O$_{10}$ (~ 0.54 nm), the measured differential conductance, and the spatial distance, respectively; The parameters $\varphi$ and $A$, with different subscripts, denote the initial phases and amplitudes of different components, while $C_1$ and $C_2$ are coefficients for a linear background, which has minimal impact on our analysis. The fitted results are shown as the black dashed curves in Fig. 3(e), and the primary fitting parameters are listed in Table I. It is obvious that the relative proportions of four components vary with energy, resulting in complex energy-dependent DOS oscillation patterns and periods in real space [Fig. 2(a) and Fig. S2 of SM]. For $P_1$ and $P_2$, the dominant periods of DOS oscillations are both 1.316$b$; for $P_3$ and $P_4$, the DOS primarily oscillates with periods of 2$b$ and 4$b$; while for $P_5$, all four components contribute equally, resulting in a more complex pattern. When focusing solely on the CDW component, we find that the CDW phase is nearly reversed between $P_3$ and $P_4$, consistent with the typical characteristics of CDW [48,49]. Additionally, $P_3$ and $P_4$ correspond to the energies where rapid DOS depletion begins (Fig. 3(b)), which are identified as the edges of a CDW gap. Therefore, our study suggests a CDW gap between -32 meV to +39 meV, i.e. 2$\Delta \approx$ 71 meV.

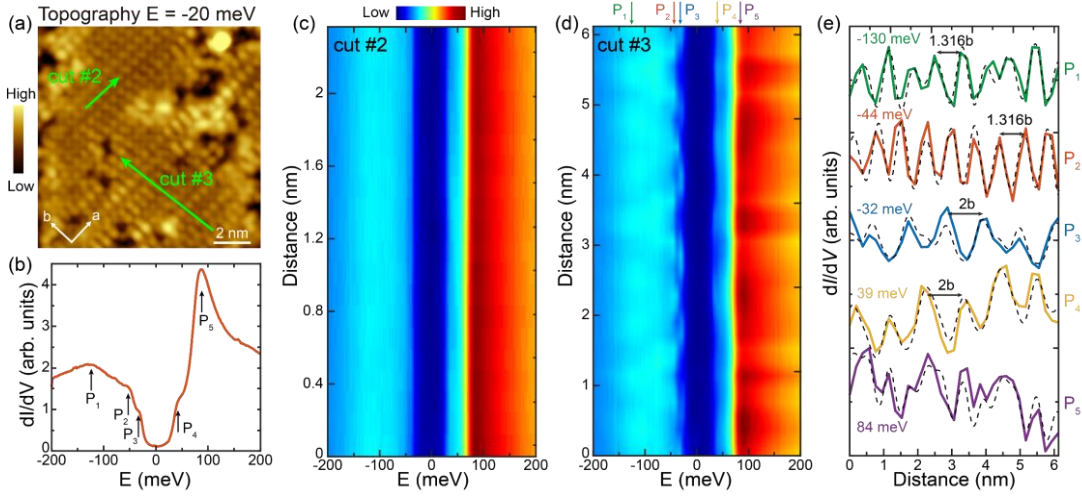

FIG. 3. Typical d$I$/d$V$ spectra and DOS oscillations on LaO-I surface. (a) Typical topographic image of LaO-I surface. (b) Typical d$I$/d$V$ spectrum of LaO-I surface, exhibiting several peaks labeled as $P_1$ - $P_5$, respectively. (c),(d) Color plots of d$I$/d$V$ spectra collected along cuts #2 and #3 in panel (a). (e) Spatial DOS oscillations at energies corresponding to $P_1 - P_5$ along cut #3, which are shifted and scaled for clarity. The fitting results are plotted out by the broken lines. Measurement conditions: (a) $V_b$ = -20 mV, $I_t$ = 30 pA; (b)-(d) $V_b$ = -200 mV, $I_t$ = 50 pA, $\Delta V$ = 4 mV.

**TABLE I. Fitting parameters for the DOS oscillations of $P_1$-$P_5$.**

| Peak | $A_b$ | $A_{CDW}$ | $A_{2b}$ | $A_{4b}$ | $\varphi_{CDW}$ |
|---|---|---|---|---|---|
| $P_1$ | 0.0346 | 0.1805 | 0.0958 | 0.0324 | $1.25\pi$ |
| $P_2$ | 0.0730 | 0.6443 | 0.0313 | 0.1475 | $0.33\pi$ |
| $P_3$ | 0.1916 | 0.1487 | 0.7429 | 0.4857 | $0.41\pi$ |
| $P_4$ | 0.0504 | 0.0653 | 0.4029 | 0.2407 | $1.46\pi$ |
| $P_5$ | 0.2644 | 0.2899 | 0.2768 | 0.4555 | $1.28\pi$ |

Although a DW-like transition has been suggested in La$_4$Ni$_3$O$_{10}$ by several techniques [24-30,35-38,42], our study provides the first direct visualization of CDW in real space. The CDW exhibits several key characteristics: 1) it is incommensurate with $\mathbf{q}_{CDW} \approx 0.76\mathbf{q}_b$ and propagates unidirectionally along $b$-axis coincided with the tilt of NiO$_6$ octahedra; 2) it is robust against surface lattice reconstructions; 3) its gap size of $2\Delta \approx 71$ meV aligns well with $T_{DW}$ = 138 K. These observations suggest an electronic origin of the CDW, such as nesting of Fermi surface patches [50,51]. We then consider possible nesting conditions based on the experimental Fermi surface (FS) of La$_4$Ni$_3$O$_{10}$ by ARPES measurements [36], as sketched in Fig. 4. Although La$_4$Ni$_3$O$_{10}$ possesses an orthogonal lattice in $ab$ plane, and the surface is further distorted after cleavage, the measured FS by ARPES is rather C$_4$-symmetric [36]. Given that the CDW is unidirectional along $b$-axis, we focus solely on the nested wave vectors along the $\overline{\Gamma}$ - $\overline{Y}$ direction.

We initially attempted to use $\mathbf{q}_{CDW}$ and 1 - $\mathbf{q}_{CDW}$ for nesting; however, there are no two parallel FS patches that could be connected by these wave vectors. Nevertheless, previous XRD and neutron diffraction studies have demonstrated that CDW and SDW emerge simultaneously below $T_{DW}$, with the wave vector $\mathbf{q}_{CDW} = 2\mathbf{q}_{SDW}$ [27]. Similar relationships have been observed in metal Cr, MnP, and (Li,Fe)OHFeSe, indicating intertwined CDW and SDW orders, where the SDW typically dominates [51-53]. Xu *et al.* did not observe any signature of CDW amplitude mode in $La_4Ni_3O_{10}$, demonstrating that SDW is more predominant [37]. Considering this possibility, we use $\mathbf{q}_{SDW}$ = 0.38$\mathbf{q}_b$ and 1 - $\mathbf{q}_{SDW}$ = 0.62$\mathbf{q}_b$ for nesting. Two possible nesting conditions are illustrated in Fig. 4. The magenta arrow connects two parallel FS patches of β bands near $\bar{S}$ point of Brillouin zone, with a wave vector of ~ 0.38$\mathbf{q}_b$, very close to $\mathbf{q}_{SDW}$; while the red arrow connects the parallel FS patches of α band at $\bar{\Gamma}$ and β band at $\bar{Y}$, with a wave vector of ~ 0.6$\mathbf{q}_b$, close to 1 - $\mathbf{q}_{SDW}$. It is worth noting that the latter aligns with previous theoretical calculations that the susceptibility reaches maxima at this wave vector [27]. These analyses support the idea that a SDW is induced by FS nesting in $La_4Ni_3O_{10}$, with the observed CDW as an accompanying order. Moreover, the unidirectionality of SDW/CDW might be related to in-plane bond anisotropy, probably the unidirectional tilt of $NiO_6$ octahedra along *b*-axis, as studied before in cuprates [54,55].

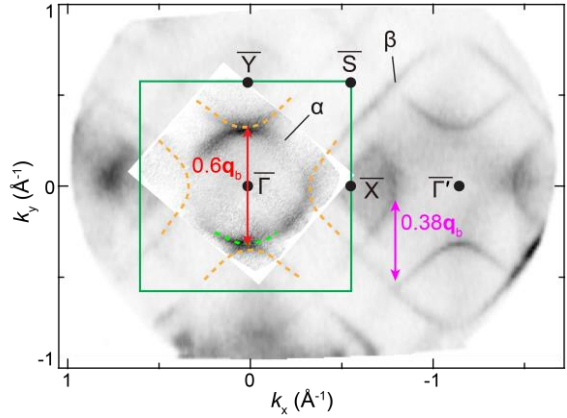

FIG. 4. Possible nesting scenarios for the DWs in $La_4Ni_3O_{10}$. The experimental FS is reproduced from ref. [36].

CDWs typically open an energy gap on the FS patches connected by their wave vectors, thus determining the gap's location on the FS is a direct way to identify the correct nesting scenario. Our STM study suggests a DW gap of 2Δ ≈ 71 meV, but its precise location on the FS is difficult to determine because of lacking of momentum resolution in STM. We attempt to compare our results with those from other techniques, but encounter significant discrepancies in both gap size and corresponding momentum locations. Optical spectroscopy and NMR measurements have revealed a DW gap of 60 meV and 50 meV, respectively [37,38]; while APRES measurements by two groups have reported an energy gap of 12 meV and 20 meV, which locate at different FS patches [25,36]. The discrepancies between different techniques highlight the need for further investigation.

The incommensurate SDW induced by FS nesting in $La_4Ni_3O_{10}$ suggests an itinerate magnetism picture. An SDW-like transition around 150 K has also been proposed in $La_3Ni_2O_7$, but with a commensurate wave vector and spin stripe orders based on an effective Heisenberg model as proposed in Refs. [32,33], which differ from $La_4Ni_3O_{10}$. Although the FS configurations of $La_3Ni_2O_7$ and $La_4Ni_3O_{10}$ are very similar [25,36,56-58], the differences in SDW properties suggest distinct magnetic exchange interactions. Optical spectroscopy and ARPES studies have revealed that the electronic correlations in $La_3Ni_2O_7$ are much stronger than in $La_4Ni_3O_{10}$ [25,35-37,40,56-59], with the former possibly favoring local magnetic exchange interactions [32,33]. This difference in electronic correlation and magnetic interaction may help explain the significant variation in $T_c$ values between pressurized $La_3Ni_2O_7$ and $La_4Ni_3O_{10}$.

Furthermore, as revealed in the phase diagram [8-11], superconductivity emerges when the DWs are suppressed at high pressures, indicating a competitive relationship between DWs and superconductivity. Both our STM results and previous optical spectroscopy experiments show severe DOS depletion at $E_F$ in $La_4Ni_3O_{10}$ below $T_{DW}$, which is unfavorable for the formation of superconductivity at ambient pressure. To further compare the properties of $La_3Ni_2O_7$ and $La_4Ni_3O_{10}$, STM studies on $La_3Ni_2O_7$ are also urgent, as it can directly reveal the microscopic details of spin and charge modulations. We are aware of a recent STM work on $La_3Ni_2O_7$ that reported gaplike features within +98 meV and -92 meV, but the rough surface limits the observation of DOS modulation [60].

In summary, we report the direct visualization of an incommensurate unidirectional CDW in $La_4Ni_3O_{10}$ in real space, and reveal a DW gap of $2\Delta \approx 71$ meV accompanied by significant DOS depletion near $E_F$. Possible FS nesting scenarios are proposed, suggesting that an SDW with $\mathbf{q}_{SDW} = 1/2\mathbf{q}_{CDW}$ is the parent phase of the observed CDW. We also compare our findings with those from other techniques and discuss the differences between $La_3Ni_2O_7$ and $La_4Ni_3O_{10}$, suggesting that the weaker electronic correlation in $La_4Ni_3O_{10}$ may be one reason for the lower $T_c$.

## ACKNOWLEDGMENTS

We thank Prof. L. X. Yang for helpful discussion. This work is supported by the National Natural Science Foundation of China (Grants No. 12494593, No. 12374140, No. 11790312, No. 12004056, No. 11774060, No. 92065201, No. 12234006), the National Key R&D Program of the MOST of China (Grants No. 2023YFA1406304, No. 2022YFA1403202), the Innovation Program for Quantum Science and Technology (Grant No. 2021ZD0302803, No. 2024ZD0300103), and the New Cornerstone Science Foundation.